\documentclass[12pt]{iopart}

\expandafter\let\csname equation*\endcsname\relax
\expandafter\let\csname endequation*\endcsname\relax
\usepackage{amsmath}
\usepackage{amssymb}
\usepackage{iopams}
\usepackage{bm}
\usepackage{color}
\usepackage{soul}
\usepackage{graphicx}

\newcommand{\abs}[1]{\lvert #1 \rvert}

\newcommand{\beq}{\begin{equation}}
\newcommand{\eeq}{\end{equation}}
\newcommand{\bea}{\begin{eqnarray}}
\newcommand{\eea}{\end{eqnarray}}

\newcommand{\eF}{\varepsilon_{F}}

\newcommand{\kF}{k_F}

\newcommand{\din}{d_{\textrm{i}}}
\newcommand{\dout}{d_{\textrm{f}}}

\newcommand{\I}{\mathrm{i}}

\providecommand{\exclude}[1]{}

\begin{document}

\title{Quantum vortex dipole as a probe of the normal component distribution}

\author{Andrea Barresi${}^{1}$, Piotr Magierski${}^{1,2}$, Gabriel Wlaz\l{}owski${}^{1,2,*}$}

\address{$^1$ Faculty of Physics, Warsaw University of Technology, Ulica Koszykowa 75, 00-662 Warsaw, Poland}
\address{$^2$ Department of Physics, University of Washington, Seattle, Washington 98195--1560, USA}
\address{$^*$ Corresponding author}
\ead{gabriel.wlazlowski@pw.edu.pl}

\begin{abstract}
We investigate the dynamics of quantum vortex dipoles in a strongly interacting, spin-imbalanced Fermi superfluid at zero temperature. Using fully microscopic time-dependent density functional theory, we demonstrate that the dipole trajectory is strongly influenced by the spatial distribution of spin polarization. The resulting forces on the vortices include both longitudinal and transverse components, leading to deflection and shrinking of the dipole during propagation. For moderate polarization, vortex dipoles are deflected and lose energy, while for larger imbalances, they are rapidly annihilated. Our findings provide compelling evidence that spin-imbalanced Fermi gases contain a spatially nonuniform normal component even at zero temperature. We show that vortex dipoles serve as sensitive probes of this component, offering a route to indirectly detect exotic superfluid phases such as the Fulde–Ferrell–Larkin–Ovchinnikov state and related inhomogeneous condensates.
\end{abstract}



\section{Introduction}

\hspace{1.5em} Quantized vortices represent the macroscopic manifestation of the quantum nature of superfluids~\cite{Donnelly1991, PethickSmith2002, Fetter2009}. They are an immediate consequence of the notion of the condensate wave function, which also gives rise 
to vorticity quantization. However, the vortex dynamics can be described in terms of classical equations of motion if the forces acting on it are known~\cite{Schwarz1985, RevModPhys.59.87}. In this picture, 2D vortex moves as a particle according to  Newton's law.   
The forces acting on it arise from the interaction between the vortex and the superfluid medium and its elementary excitations, and were a subject of considerable theoretical and experimental efforts, see~\cite{Barenghi1983Aug, Sergeev2023} for reviews.
In particular, the dissipative forces, which originate from interaction with the normal component of the superfluid, are still debatable.
The first attempt to determine the forces due to phonon scattering in a bosonic superfluid was made by Iordanskii \cite{iordanskii1,iordanskii2,iordanskii3}, who showed that it generates a transverse component of the effective force.
While there is an established consensus that the Iordanskii force represents the interaction between a vortex and the quasiparticles in the superfluid, theories addressing its magnitude and importance in vortex dynamics diverge.
There are arguments that it is effectively zero~\cite{aothouless1,aothouless2,demircan}, and claims that it is a non-negligible component of the effective Magnus force~\cite{sonin,kopnin1976,kopnin1995,wexler,geurst}. There are also attempts to synthesize both points of view,  making its magnitude dependent on the regime of velocities~\cite{gavassino}. Unfortunately,
due to the lack of experimental evidence,  no definitive conclusion has been 
reached (see \cite{magierski2024} for a review). The difficulties arise from attempts to generalize the Magnus force in the presence of both components: normal and superfluid. Typically, the superfluid component reflects the presence of the condensate, while the normal component is connected to the presence of thermal excitations. However, there are possible situations in which the normal component may be present even at zero temperature. One example is when the system is characterized by density modulations, which break Galilean symmetry, as shown by A. Leggett~\cite{PhysRevLett.25.1543,Leggett1998}, and has recently been the active area of research in the context of ultracold gases~\cite{PhysRevLett.130.226003,orso2023,perezcruz2024} and neutron stars~\cite{almirante2025}. The other possible scenario, which is relevant to this paper, requires the consideration of superfluid Fermi systems with spin 
imbalance. Superfluidity in Fermi systems arises because of the creation of a condensate of Cooper pairs. But, since there is an unequal number of spin-up and spin-down particles, we expect some of them to be unpaired, which effectively should create a normal fraction. It is also known  
that the presence of spin imbalance has an impact on the structure of 
the vortex core~\cite{magierskispin,magierski2024}. As a consequence, the effects 
related to the scattering of the normal component off the vortex core are expected
to be significantly affected.

The phase diagram of spin-imbalanced systems is still not well established. Various scenarios were proposed; see reviews~\cite{Radzihovsky2010,Chevy2010,Gubbels2013}. The one that is discussed most frequently was proposed by Larkin and Ovchinnikov (LO)~\cite{lo} and Fulde and Ferrel (FF)~\cite{ff}. The characteristic property of such states is the spatially modulated order parameter, such as $\Delta(\bm{r})\sim|\Delta|e^{i\bm{q}\cdot\bm{r}}$ for the FF-type state and $\Delta(\bm{r})\sim|\Delta|\cos(\bm{q}\cdot\bm{r})$ for the LO-like state, where the modulation wave vector $\bm{q}$ is set by the difference of the Fermi wave vectors of the individual spin components. Recent numerical simulations~\cite{ferron-fflo} point to the conclusion that the ground state of spin-imbalanced Fermi gases may exhibit a variety of possibilities depending on the amount of the spin-imbalance, ranging from randomly distributed spin-polarized droplets (also called ferrons~\cite{ferron1} or ring solitons~\cite{PhysRevResearch.2.043282}), via disordered structures similar to liquid crystals, up to periodically modulated states like LO. The common property of all these scenarios is the position dependence of the order parameter $\Delta(\bm{r})$, which develops spontaneously.
It implies that the majority spin particles are accumulated in regions where $\Delta$ vanishes. 
Consequently, the normal component is distributed non-uniformly across the spin-imbalanced system. 
Unfortunately, such a picture has not been confirmed experimentally so far, mainly due to a lack of methods for directly observing the distribution of the order parameter or the normal component; see~\cite{Kinnunen2018Feb} for a recent review of attempts to detect the LOFF state in ultracold Fermi gases. 

\begin{figure}[t]
	\centering
	\includegraphics[width=0.5\textwidth]{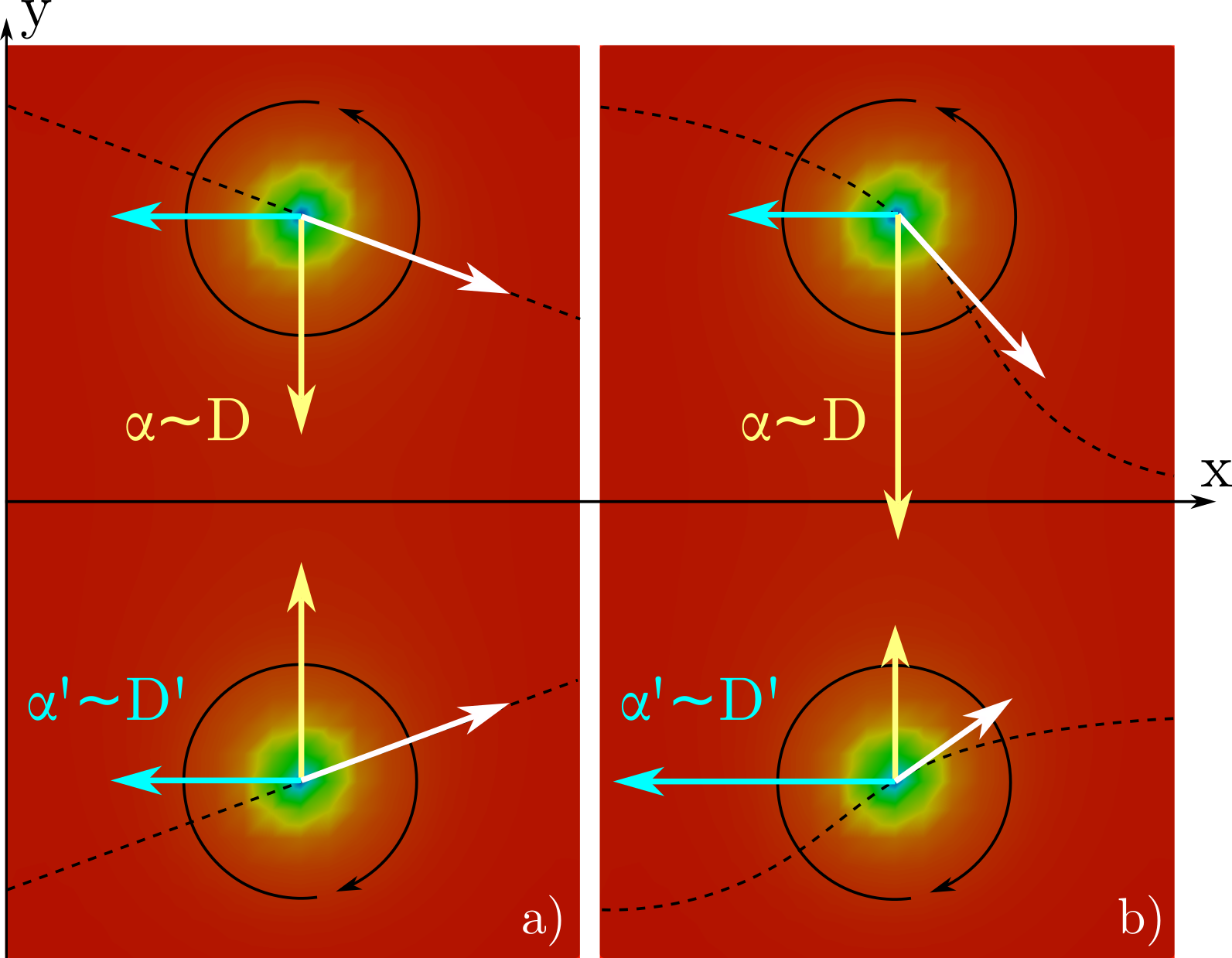}
	\caption{Comparison of forces acting on vortex cores under different density regimes. a)~Both normal and superfluid component are isotropic ($\rho_s = \textrm{const}, \rho_n = \textrm{const}$); b)~Both components are distributed inhomogeneously ($\rho_s = \rho_s(\bm{r}), \rho_n = \rho_n(\bm{r})$), but their sum is isotropic $\rho_s(\bm{r})+\rho_n(\bm{r})\approx \textrm{const}$.}
	\label{fig:forces}
\end{figure}
Advances in manipulating and detecting quantum vortices in superfluid gases allow us to use them as probes of properties of the superfluid medium. They were already widely used in the context of studies of Bose-Einstein condensate properties; see the review articles~\cite{Fetter2010,Verhelst2017}. 
Recently, experimental techniques reached a level that allows the creation of vortices one by one at will and, subsequently, to control and image them with high accuracy. For example, in experiment~\cite{Kwon2021Dec} by the LENS group, vortex dipoles were collided to gain insight into dissipative processes in fermionic superfluids. Due to its structural simplicity on one side and the ability for controlled creation on the other, the vortex dipole emerges as a very tempting probe for studies of fundamental properties of superfluid states~\cite{PhysRevLett.104.160401,PhysRevA.93.023603,PhysRevA.93.023604,grani2025}.
Consequently, taking into account these two factors: the modification of the vortex core structure and the enhancement of the normal component in the presence of spin imbalance, one expects
significant modification of the vortex dipole dynamics in such conditions, even at $T=0$.

Many open questions remain about the role of the normal component in strongly interacting Fermi systems, particularly at zero temperature and under the spin-imbalance. Spin-polarized Fermi gases are predicted to host exotic superfluid phases with spontaneously modulated order parameters, but experimental confirmation of these states has been elusive due to the absence of direct probes of the normal fraction. Motivated by recent advances in vortex manipulation and imaging, this work addresses two key questions: (i) How does the presence of spin imbalance and associated spatial inhomogeneities influence vortex dipole dynamics in strongly interacting Fermi superfluids? (ii) Can vortex dipole trajectories provide quantitative insight into the distribution and strength of the normal component, and thus serve as a diagnostic for exotic pairing states?

In this work, we investigate the dynamics of a quantum vortex dipole in a spin-imbalanced and strongly interacting Fermi superfluid. We use a microscopic approach in the form of Time-Dependent Density Functional Theory (TDDFT) extended to superfluid systems~\cite{Bulgac2007,Bulgac2012} with nonzero spin polarization~\cite{PhysRevLett.101.215301}. The approach is free from any assumptions regarding the relative interaction between quantum vortices and components of the system. From the obtained vortex dipoles trajectories, we demonstrate unambiguously that spin-imbalanced Fermi superfluids are characterized by the presence of the normal component even at zero temperature, which is distributed inhomogeneously across the gas. The work provides a practical procedure for detecting exotic phases in spin-imbalanced superfluid Fermi gases.

The remainder of this paper is organized as follows. Sec.~2 introduces the theoretical framework for describing vortex dipole motion in terms of effective forces and establishes its connection to the presence of the normal component. Sec.~3 describes the numerical approach based on time-dependent density functional theory for spin-imbalanced Fermi gases. Sec.~4 presents simulation results, analyzing dipole trajectories across varying levels of polarization and their interpretation within a point-vortex model. Finally, Sec.~5 summarizes the main findings, discusses their implications for experimental detection of inhomogeneous superfluid phases, and outlines future research directions.

\section{Vortex dipole as a probe}
Let us consider the propagation of the vortex dipole (a vortex-antivortex pair) in a superfluid medium in the presence of a normal component. From the perspective of the effective vortex point model~\cite{Eyink2006,Richaud2020,Richaud2021,Skipp2023}, the dynamics of $i$-th vortex ($i=1,2$) is governed by Newton's equation
\begin{equation}
m_i\ddot{\bm{r}}_i=\bm{F}_M+\bm{F}_N.
\end{equation}
where $m_i$ denotes the vortex mass. In most treatments, vortices are regarded as massless objects ($m_i\approx 0$). In this limit, the above equation reduces to the first-order force-balance condition
\begin{equation}
\bm{F}_M+\bm{F}_N\approx 0,
\end{equation}
since both forces depend explicitly on the vortex velocity $\bm{v}_i = \dot{\bm{r}}_i$. Namely
\begin{equation}
\bm{F}_M=\rho_s\kappa\bm{\hat{z}}\times (\bm{v}_i - \bm{v}_s^{(j\neq i)})
\label{eq:FM}
\end{equation}
is Magnus force which arise from superflow $\bm{v}_s^{(j\neq i)}$ generated by other vortex, and
\begin{equation}
\bm{F}_N=D(\bm{v}_i - \bm{v}_n) + D^{\prime}\bm{\hat{z}}\times(\bm{v}_i - \bm{v}_n)
\label{eq:FN}
\end{equation}
is the frictional force. It has two components, longitudinal ($\sim D$) and transverse ($\sim D^{\prime}$), which in the case of vortex dipole induce two effects: affecting the speed of the dipole propagation 
and decreasing the relative distance between vortices, see Fig.~\ref{fig:forces}(a).
This can be clearly seen if one explicitly writes the expression for vortex dipole velocities originating from the force balance. Namely, for the vortex-antivortex pair, with circulation $\kappa$ and separated by the distance $d(t)$:
\begin{eqnarray}\label{velocities}
\bm{v}_{1}(t) &=&\frac{|\kappa|}{d(t)} \left [ (1 - \alpha^\prime)\bm{\hat{y}} - 
            |\alpha|\bm{\hat{x}} \right ], \nonumber \\
\bm{v}_{2}(t) &=& \frac{|\kappa|}{d(t)} \left [ (1 - \alpha^\prime)\bm{\hat{y}} + 
            |\alpha|\bm{\hat{x}} \right ],
\end{eqnarray}
where $\alpha=\frac{\tilde{D}}{\tilde{D}^2+(1-\tilde{D}^\prime)^2}$, 
$1-\alpha^\prime=\frac{1-\tilde{D}^\prime}{\tilde{D}^2+(1-\tilde{D}^\prime)^2}$, 
$\tilde{D}=\frac{D}{\kappa\rho_s}$, $\tilde{D}^\prime=\frac{D^\prime}{\kappa\rho_s}$. 
The reference frame in these equations is defined in the following way:
for an arbitrary vortex, direction of $\bm{\hat{z}}$ is specified by its vorticity, then $\bm{\hat{x}}$ is the perpendicular vector pointing towards the other vortex. Consequently $\bm{\hat{y}} = \bm{\hat{z}}\times\bm{\hat{x}} $ points
along the direction of motion of the dipole. 
Note that the form of the equations~(\ref{velocities}) does not depend on the choice of the initial vortex, and the behavior of vortices is symmetric with respect to $\bm{\hat{x}}$ direction.
The equations ($\ref{velocities}$) describe two effects related to the dissipative forces: the slowing down of the vortex dipole, which is governed by the coefficient $\alpha^\prime$, and the decrease in the intervortex distance $d$, which is governed by the coefficient $\alpha$.
Consequently, the energy of the dipole decreases, since it is related to its size $E_{\textrm{dipole}}\sim \ln(d/\xi)$, where $\xi$ is the coherence length~\cite{Donnelly1991}. 

In this work, we focus on fermionic vortices. A key difference between vortices in Bose and Fermi superfluids lies in their internal structure. While bosonic vortices typically have empty cores at zero temperature, fermionic vortices host discrete Caroli–de Gennes–Matricon (CdGM) states. These states provide additional channels for dissipative processes via quasiparticle scattering~\cite{Kopnin2002,silaev2012universal} and dominate the low-temperature contribution to the Iordanskii force. Moreover, the matter contained within fermionic vortex cores is expected to give rise to a nonzero vortex mass~\cite{richaud2024}. As a result, the coupling of vortices to the normal component, expressed in the point-vortex model by the $D$ and $D^\prime$ constants, is generally stronger than in the bosonic case, making fermionic vortices more sensitive probes of the normal component in the system. Recent measurements of the dissipation coefficients in the unitary Fermi gas have confirmed that CdGM states contribute significantly to their values~\cite{grani2025}.

In the case of a spin-imbalanced system, which induces inhomogeneity, the friction coefficients $D$ and $D^{\prime}$ may exhibit spatial variations. They originate from the variation of the superfluid-to-normal component of the total density. Although it is not clear what the size of spatial modulation is, it is reasonable to assume that the main contribution to the dissipative forces comes from the neighborhood of the vortex core~\cite{Sergeev2023}. For example, difficulty in determining $D^{\prime}$ lead to controversy with the Iordanskii force~\cite{aothouless2,RevModPhys.59.87,Barenghi1983Aug}. Typically, coefficients standing in front of forces are regarded as functions of temperature, which quantifies the amount of the normal component. For Fermi superfluids in the deep BCS limit, the semiclassical expressions have been derived by Kopnin~\cite{Kopnin2002}
\begin{align}
    \tilde{D}=&
    \frac{\rho}{\rho_s}\frac{\omega_0\tau_{\textrm{eff}}}{\omega_0^2\tau_{\textrm{eff}}^2+1}\tanh\frac{\Delta}{2T}, & 
    \tilde{D}^\prime=& 
    1-\frac{\rho}{\rho_s}\frac{\omega_0^2\tau_{\textrm{eff}}^2}{\omega_0^2\tau_{\textrm{eff}}^2+1}\tanh\frac{\Delta}{2T},
\end{align}
where $\rho$ stands for the total density, being sum of superfluid  $\rho_s$ and normal  $\rho_n$ densities. Remaining quantities are: $\tau_{\textrm{eff}}$ - effective relaxation time, $\omega_0$ -  minigap energy that defines the lowest energy state in the vortex core~\cite{Volovik2009-lt,magierskispin}, and $T$ stands for the temperature. While the expressions for $D$ and $D^\prime$ depend on the considered systems and assumptions made, the feature that they are related to normal and superfluid densities remains valid. We may then identify generic features of the vortex dipole trajectory arising from the force balance equation. The shrinking of the dipole as it propagates is the signature of the presence of the normal component; if the system is purely superfluid, $d$ remains constant. 
Next, if the superfluid and normal densities are uniformly distributed, then the dissipative forces affect both vortices with the same strength, see Fig.~\ref{fig:forces}(a).
The dipole's position $\bm{r}_{\textrm{d}}=(\bm{r}_1+\bm{r}_2)/2$ moves along the straight line. 
If the normal $\rho_n$ and superfluid $\rho_s$ components exhibit position dependence, the resulting trajectory of the propagating dipole will be affected significantly, even in the uniform system ($\rho_s+\rho_n=\rho=\textrm{const}$ where $\rho$ is total density). 
The dipole no longer propagates along a straight line, and the precise form of the trajectory will depend on the distribution of the components, see Fig.~\ref{fig:forces}(b). This way, vortex dipoles can be used as robust probes if the normal component is present and, if yes, if it is distributed uniformly or nonuniformly in the system.   

\section{Numerical framework}
The superfluid TDDFT formalism, at the formal level, is equivalent to time-dependent Bogoliubov-de Gennes equations~\cite{Zhu2016} (we use units where $m=\hbar=1$)
\begin{gather}
  i\frac{\partial}{\partial t}
  \begin{pmatrix}
    u_n(\bm{r}, t)\\
    v_n(\bm{r}, t)
  \end{pmatrix}
  =
  \begin{pmatrix}
    h_{\uparrow}(\bm{r}, t)-\mu_{\uparrow} & \Delta(\bm{r}, t)\\
    \Delta^*(\bm{r}, t) & -h_{\downarrow}^*(\bm{r}, t)+\mu_{\downarrow}
  \end{pmatrix}
  \begin{pmatrix}
    u_n(\bm{r}, t)\\
    v_n(\bm{r}, t)
  \end{pmatrix},\label{eq:TDSLDA}
\end{gather}
describing time evolution of the Bogoliubov amplitudes $\{u_n,v_n\}$. These in turn, define time evolution of densities $n_{\sigma}$, currents $\bm{j}_{\sigma}$ and the order parameter $\Delta$
\begin{subequations}
  \label{eq:Densities}
  \begin{gather}
    \begin{aligned}
      \rho_{\uparrow}(\bm{r},t) &= \sum_{E_n>0}\abs{u_{n}(\bm{r},t)}^2, &
      \rho_{\downarrow}(\bm{r},t) &= \sum_{E_n>0}\abs{v_{n}(\bm{r},t)}^2,
    \end{aligned}\\
    \begin{aligned}
      \label{eq:nu_dens}
      \bm{j}_{\uparrow}(\bm{r},t) =
      -&\frac{\I}{2} 
      \sum_{E_n>0} \left[u^*_{n}(\bm{r},t) \nabla u_{n}(\bm{r},t) 
        - u_{n}(\bm{r},t) \nabla u^*_{n}(\bm{r},t)\right]
      ,\\
      \bm{j}_{\downarrow}(\bm{r},t) =&
      \frac{\I}{2} 
      \sum_{E_n>0} \left[v^*_{n}(\bm{r},t) \nabla v_{n}(\bm{r},t)  
        - v_{n}(\bm{r},t) \nabla v^*_{n}(\bm{r},t)\right]
      ,
    \end{aligned}\\
    \begin{aligned}
    \Delta(\bm{r},t) &= -\frac{\gamma}{\rho_{\uparrow}+\rho_{\downarrow}}\sum_{E_n>0} u_{n}(\bm{r},t)v_{n}^{*}(\bm{r},t).
    \end{aligned}
  \end{gather}
\end{subequations}
The sums include states only up to the cut-off energy $E_c$, and the pairing coupling constant $\gamma$ is renormalized according to the prescription given in~\cite{Bulgac2012}. The single-particle Hamiltonian contains kinetic, mean-field, and external potential terms
\begin{equation}
h_{\sigma} = \frac{1}{2}\bm{\nabla}^2 + U_{\sigma}(\bm{r},t) + V_{\sigma}^{(\textrm{ext})}(\bm{r},t),
\end{equation}
where $U_{\sigma}$ depends again on the densities. The explicit value of the coupling constant $\gamma$ and form of the $U_{\sigma}$ is given in Ref.~\cite{Bulgac2012}, where a variant of density functional theory for studies of spin-imbalanced unitary Fermi gas was presented, known as Asymmetric Superfluid Local Density Approximation (ASLDA)~\cite{PhysRevLett.101.215301}. Here we used its simplified form, where we assume that the effective mass of a particle is equal to its bare mass (i.e., parameter $\alpha_{\sigma}=1$ according to the notation in~\cite{Bulgac2012}). The chemical potentials $\mu_{\sigma}$ are used to control particle numbers of each components $N_{\sigma}=\int \rho_{\sigma}(\bm{r})d\bm{r}$, and thus the total spin imbalanced of the system
\begin{equation}
P = \frac{N_{\uparrow}-N_{\downarrow}}{N_{\uparrow}+N_{\downarrow}}.
\end{equation}
The results presented here are for the zero temperature limit. 

\begin{figure}[t]
	\centering
	\includegraphics[width=0.5\textwidth]{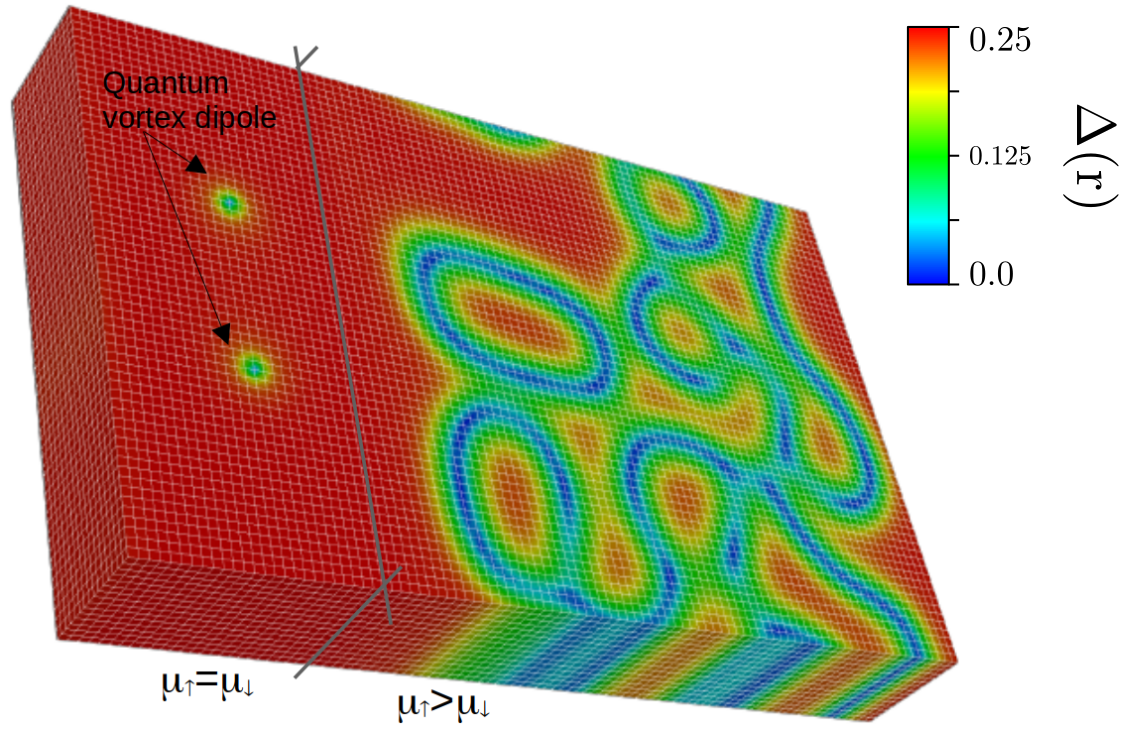}
	\caption{Example of the initial setup, represented on the spatial mesh of size $100\times 64\times 16$. The vortex dipole is imprinted in the region where the system is locally spin-balanced. During the dynamics, the dipole will propagate toward the region where the system is spin-imbalanced. The trajectory of the vortex dipole will be used as a probe for studies of the underlying structure of the spin-imbalanced environment. The color map shows the spatial distribution of the pairing field absolute value $\Delta(\bm{r})$.}
	\label{fig:setup}
\end{figure}
We solve the problem~(\ref{eq:TDSLDA}) numerically, on a spatial grid of size $N_x\times N_y\times N_z= 100\times 64\times 16$ with lattice spacings $\Delta x=\Delta y=\Delta z=1$, and with periodic boundary conditions. To simplify the computation process, we assume uniformity of the system in the third direction, and the quasiparticle wave functions acquire the form
\begin{equation}
  \begin{pmatrix}
    u_n(\bm{r}, t)\\
    v_n(\bm{r}, t)
  \end{pmatrix}
  =
    \begin{pmatrix}
    u_n(x,y, t)\\
    v_n(x,y, t)
  \end{pmatrix}e^{ik_z z},
\end{equation}
where $k_z$ are wave-vectors that take discrete values spanning the first Brillouin zone, with the step $\Delta k_z = 2\pi/N_z$. 
In the time-dependent TDDFT simulations, we monitor the conservation of both energy and particle number as our primary code's stability test. Specifically, we require relative energy variations to remain within $\Delta E(t)/E(0)\lesssim 10^{-4}$ and particle number changes within $\Delta N(t)/N(0)\lesssim 10^{-8}$.

We consider a unitary Fermi gas (UFG) with nonzero global spin polarization $P>0$. 
Such systems are expected to have inhomogeneously distributed local spin polarization 
\begin{equation}
p(\bm{r}) = \frac{\rho_{\uparrow}(\bm{r})-\rho_{\downarrow}(\bm{r})}{\rho_{\uparrow}(\bm{r})+\rho_{\downarrow}(\bm{r})}
\end{equation}
in the ground states, even in the absence of an external potential~\cite{ferron-fflo}. To investigate it, we prepare the initial state, which consists of two regions. In one, we assume that the local chemical potentials are equal (the system is locally spin-balanced). In this part, we imprint a vortex dipole, which will be used as a probe for studies of the second region in which the local chemical potentials are unequal, and the system develops various inhomogeneities.  The initial setup is obtained as the result of solving the static variant of Eq.~(\ref{eq:TDSLDA}), i.e. $i\frac{\partial}{\partial t}\rightarrow E_n$ with constraints on the chemical potentials. The vortex dipole is imposed by means of phase imprinting techniques. For more details related to the preparation of the initial states, see~\ref{Appendix:A}. An example of the initial state is shown in Fig.~\ref{fig:setup}. Since the system has translational symmetry along the third coordinate, we will be showing only the $x-y$ plane in subsequent plots. In the computation, we have used the publicly available W-SLDA Toolkit~\cite{wsldaweb}, a software package developed to solve problems formally equivalent to the fermionic Bogoliubov--de Gennes equations in both static and time-dependent formulations. In the following, we present our findings, which are largely based on~\cite{phd-barresi}. To report results in the dimensionless form, we define the Fermi wave-vector as $\kF=(3\pi^2\rho_0)^{1/3}$ with $\rho_0$ being the total density in the region where the system is locally spin-balanced, and the related Fermi energy is $\varepsilon_F = k_F^2 /2$. We consider dynamics of dipoles with the initial sizes up to $20k_F^{-1}$ so that the results are not significantly affected by interaction with the image dipoles, due to the applied periodic boundary conditions.

\section{Vortex propagation in spin imbalanced system}
\label{sec:low-p}

\begin{figure}[t]
	\centering
	\includegraphics[width=0.8\textwidth]{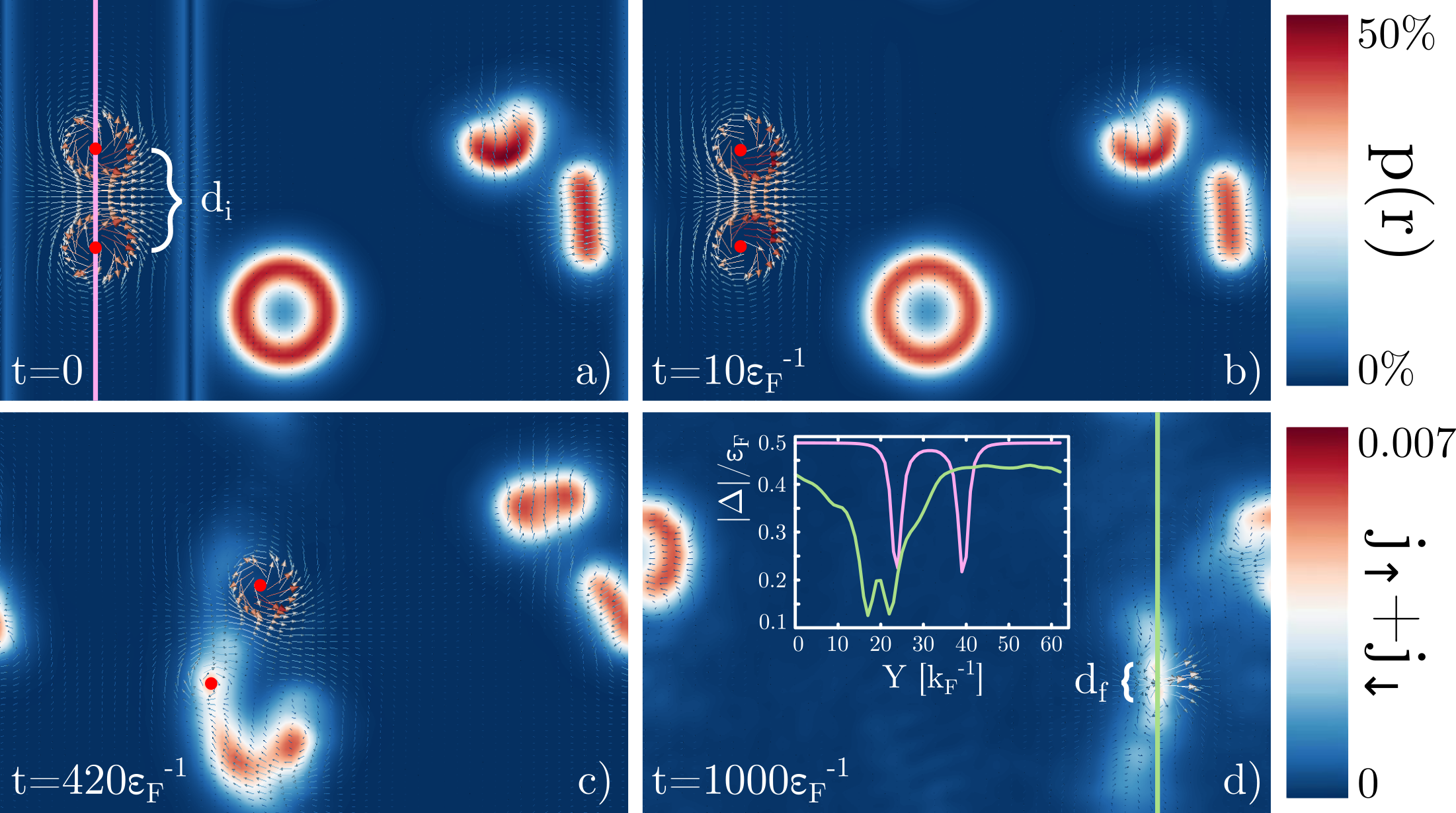}
	\caption{Example of a vortex dipole propagation in a spin-imbalanced environment with global spin-polarization $P=2.89\%$. The initial size of the vortex dipole is $d_{\textrm{i}}=16k_F^{-1}$. Panel a) shows the initial configuration. The vertical stripes visible in $p(\bm{r})$ quantity (color map) are due to the initial state preparation procedure, where we separate the simulation volume into spin-balanced and spin-imbalanced regions by means of the external potential. As the dynamics start, the system relaxes, and this artifact vanishes, as seen in panel b) for the time moment $  t \approx 10\eF^{-1}$. Panel c) $t \approx 420\eF^{-1}$: the dipole interacts with the ferron; one of the vortices sucks in the spin-polarization which significantly affects the ferron's structure. Panel d) $t \approx 1000\eF^{-1}$: the dipole size has significantly shrunk, and the cores have now been filled with the spin-polarization.
	Inset: Cross-section of the order parameter $\Delta$, along the line as shown in panels a) and d). The dipole persists, but its internal structure has significantly changed. Arrows (heat scale) indicate the local intensity and direction of the total current $\bm{j}_\uparrow(\bm{r},t)+\bm{j}_\downarrow(\bm{r},t)$. Red dots indicate positions of vortex cores. }
	\label{fig:low-pol} 
\end{figure}

We start discussions of the results for the cases of low spin imbalance $1\% \leq P \leq 5\%$. 
In this case, the number of unpaired particles ($N_{\uparrow}-N_{\downarrow}$) is small enough to allow the formation of a limited number of well-separated regions characterized by non-zero local spin polarization. 
Fig.~\ref{fig:low-pol} shows an example of this pattern. They are qualitatively similar to the structures reported in~\cite{ferron-fflo,PhysRevResearch.2.043282}, which will be referred to as ferrons herafter.
These structures form stationary configurations, i.e., they neither expand nor diffuse over time. Their stability has been analyzed in detail in Refs.~\cite{ferron1,PhysRevA.104.033304,PhysRevResearch.2.043282}, where it was demonstrated that such objects remain stable under typical conditions. In particular, Ref.~\cite{PhysRevA.104.033304} identifies a critical flow velocity above which ferrons become unstable, and shows that vortices can, in principle, destroy them. 
The connection between spin-imbalance accumulation and the length of nodal lines has also been investigated in Refs.~\cite{ferron-fflo,PhysRevA.104.033304}.

The most notable result observed during the dynamics of the probe is the modification of the dipole's trajectory as it passes the region with increased local spin imbalance. The panels (b-d) of Fig.~\ref{fig:low-pol} clearly demonstrate that the dipole is deflected towards the ferron. Moreover, its size decreases as the vortex dipole interacts with the ferron. Since the distance after the dipole-ferron interaction $d_{\textrm{f}}$ is smaller than before the interaction $d_{\textrm{i}}$, it demonstrates that the acting force has a frictional character, pointing to activation  of $F_{N}$ force given by Eq.~(\ref{eq:FN}). We also note that the vortex core structure is affected due to the interaction with the ferron, hosting unpaired particles in it after the event and exhibiting a resemblance to quantum vortices typical for spin-polarized Fermi gases~\cite{magierskispin}. 

\begin{figure*}[t]
	\centering
	\includegraphics[width=0.99\textwidth]{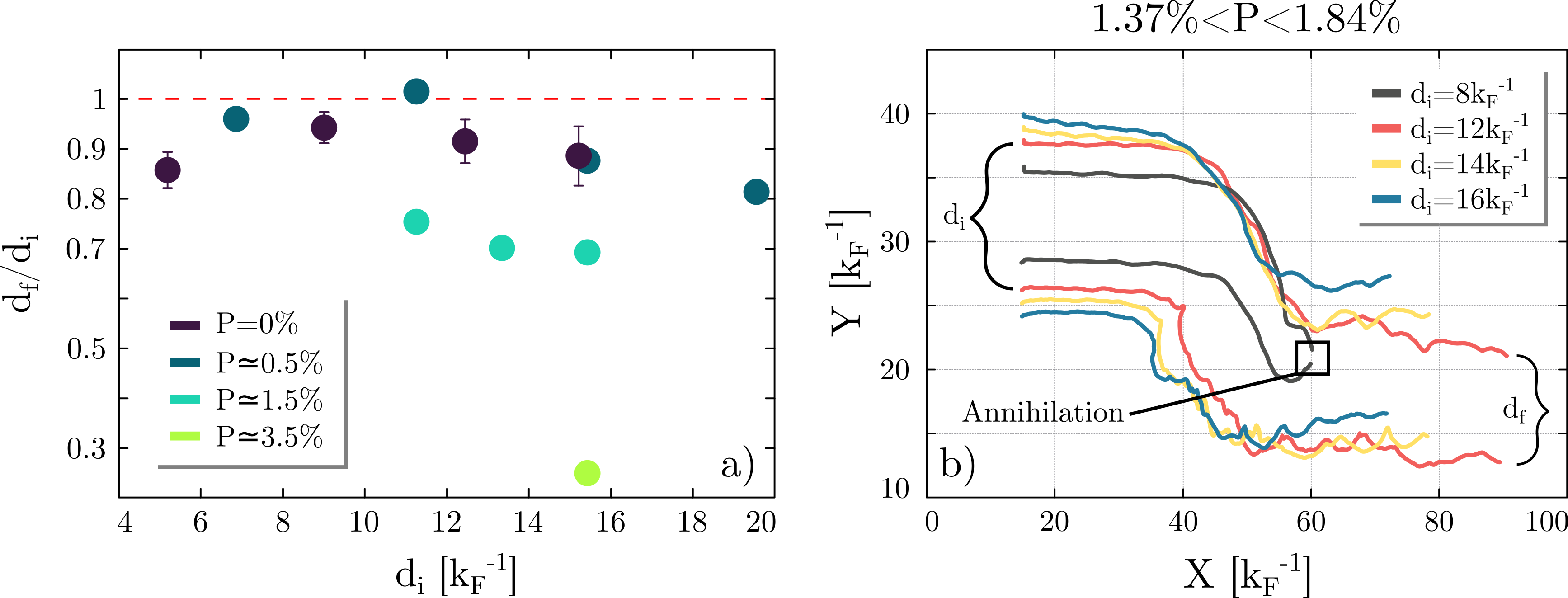}
	\caption{
		Panel a) Distance ratio $\dout/\din$ as a function of initial dipole size $d_{\textrm{i}}$ for various polarizations. Notably, the series for $P \simeq 3.5\%$ has only one datapoint for which the dipole survives. For lower $d_{\textrm{i}}$, the dipoles annihilate before the trajectory ends.
		Panel~b) Trajectories of vortex dipoles with different initial sizes $d_{\textrm{i}}$ at fixed polarization $P \approx 1.5\%$. The dipole changes the direction of propagation upon colliding with a ferron. For the smallest initial size, the dissipation is high enough that the dipole annihilates.}
	\label{fig:dfdi-lowpol}
\end{figure*}

As the vortex dipole passes close to the ferron, we observe a trajectory deflection towards the ferron position and a shrinking with respect to the initial dipole size.
The trajectories for different initial dipole sizes are shown in Fig.~\ref{fig:dfdi-lowpol}b. These were realized with the use of a detection algorithm, which works in the following way. Given an initial estimate for the coordinates of the vortex core, the algorithm scans the area within a 5 lattice units radius and identifies the minimum of $\Delta(r)$. This operation is repeated at every timestep, for which the coordinates of the vortex core are recorded.
The initial size $d_{\textrm{i}}$ defines not only the dipole energy but also the propagation velocity. This, in turn, quantifies the time interval over which a vortex interacts with the localized normal component. These factors impact the size of the deflection, but they do not significantly modify the amount of dissipated energy. For example, for $P\approx 1.5\%$, the dipole size decreases by a factor $d_{\textrm{f}}\approx 0.7 d_{\textrm{i}}$. The change in the dipole size for scenarios with different global spin polarizations $P$ is shown in Fig.~\ref{fig:dfdi-lowpol}a. Note that the size of the dipole cannot be smaller than a critical size, being of the order $\sim 2\xi$. 
If the friction with the normal component causes the vortex dipole to shrink below this threshold, the vortex and antivortex cores begin to overlap, triggering annihilation. This process passes through a transient configuration resembling the Jones–Roberts solitons~\cite{Jones1982,JRS}, which eventually decays into sound waves.
We find that for global polarizations $P \lesssim 0.5\%$, the dissipative effects are marginal. The change of $d_{\textrm{f}}/d_{\textrm{i}}$ is mainly due to the vortex acceleration. Namely, our imprinting method produces the dipole with zero initial velocity. When it speeds up, at the beginning of the simulation, a fraction of its energy is transferred into kinetic energy. This effect was pointed out in Ref.~\cite{barresi}, from which we also took data series for $P=0\%$ in Fig.~\ref{fig:dfdi-lowpol}a. For $P\gtrsim1.5\%$, we observe a significant enhancement of the dissipative dynamics, which we attribute to the induction of the normal component due to the presence of unpaired particles. Consistently, the minimum size of the dipole that is able to survive interaction with ferrons (within the studied time interval $\approx 2000\eF^{-1}$) increases as well, being about $10\kF^{-1}$ and $15\kF^{-1}$ for polarizations $P=1.5\%$ and $3.5\%$ respectively. 

\begin{figure*}[t] 
	\centering
	\includegraphics[width=0.6\textwidth]{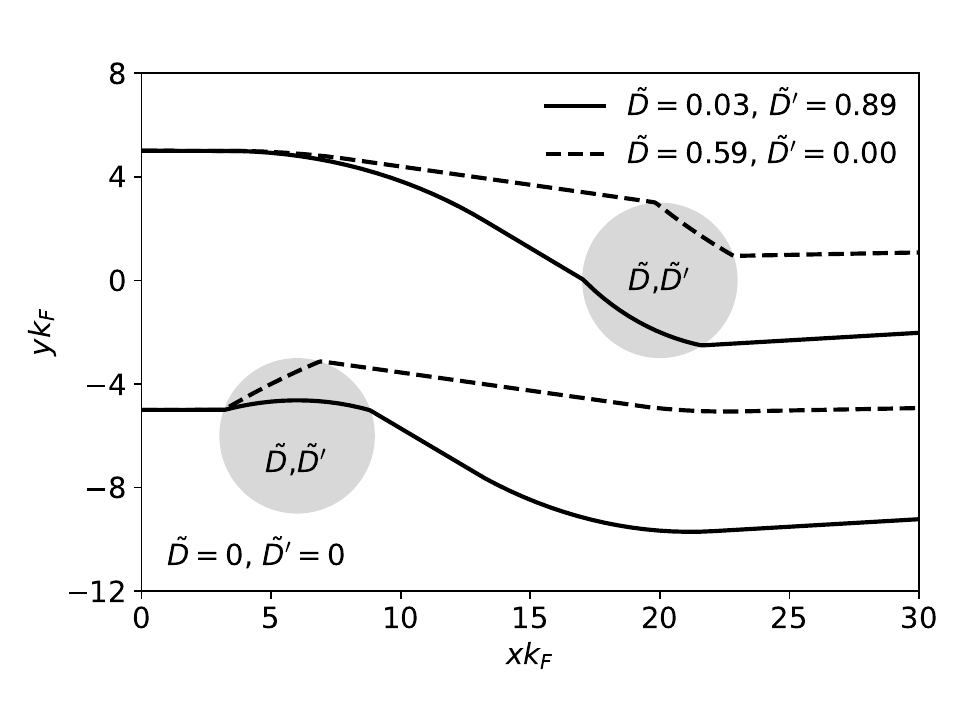}
	\caption{Solution of classical equations of motion for the massless vortex dipole, assuming that the dimensionless dissipative coefficients $\tilde{D}$ and $\tilde{D}^\prime$ acquire nonzero values only in selected regions, indicated as grey circles. Two cases are presented: the transverse force dominates over the longitudinal (solid line) and the transverse force is absent (dashed line).}
	\label{fig:pvm}
\end{figure*}
The deflection of the dipole trajectory clearly indicates that the forces acting on the vortex and antivortex differ in magnitude. This asymmetry can only arise if the coefficients, such as $D$ and $D^\prime$, depend on position. To demonstrate this, we numerically solved the classical equations of motion, $\bm{F}_M+\bm{F}_N=0$, for a vortex dipole, assuming that the normal component is localized exclusively within ferrons. This effectively means that $D$ and $D^\prime$ are nonzero only in those regions. 
In general, TDDFT simulations reveal that ferrons can move, for instance as a result of interactions with vortices. This behavior could also be incorporated into the classical framework by introducing equations of motion for the ferrons. However, for the sake of simplicity, in the present tests we restrict ourselves to a minimal model in which ferrons are assumed to remain stationary ($\bm{v}_n=0$). Example solutions within this approximation are shown in Fig.~\ref{fig:pvm}.
Remarkably, even this minimal model reproduces the deflected trajectories observed in microscopic simulations. We find that the longitudinal force mainly affects the dipole size, while the transverse force typically causes the trajectory to bend. This highlights the crucial role of the transverse component in vortex dynamics within the Fermi superfluid. These findings align with recent measurements of dissipative coefficients in the unitary Fermi gas, which showed that the transverse coefficient $\alpha^\prime$ (proportional to $D^\prime$) dominates over the longitudinal coefficient $\alpha$ (proportional to $D$)~\cite{grani2025}. Nonetheless, we cannot entirely rule out the possibility that a purely longitudinal force could also reproduce the observed behavior, given an appropriately tuned distribution of the normal component and without assuming its stationarity.

\begin{figure}[t]
	\centering
	\includegraphics[width=0.8\columnwidth]{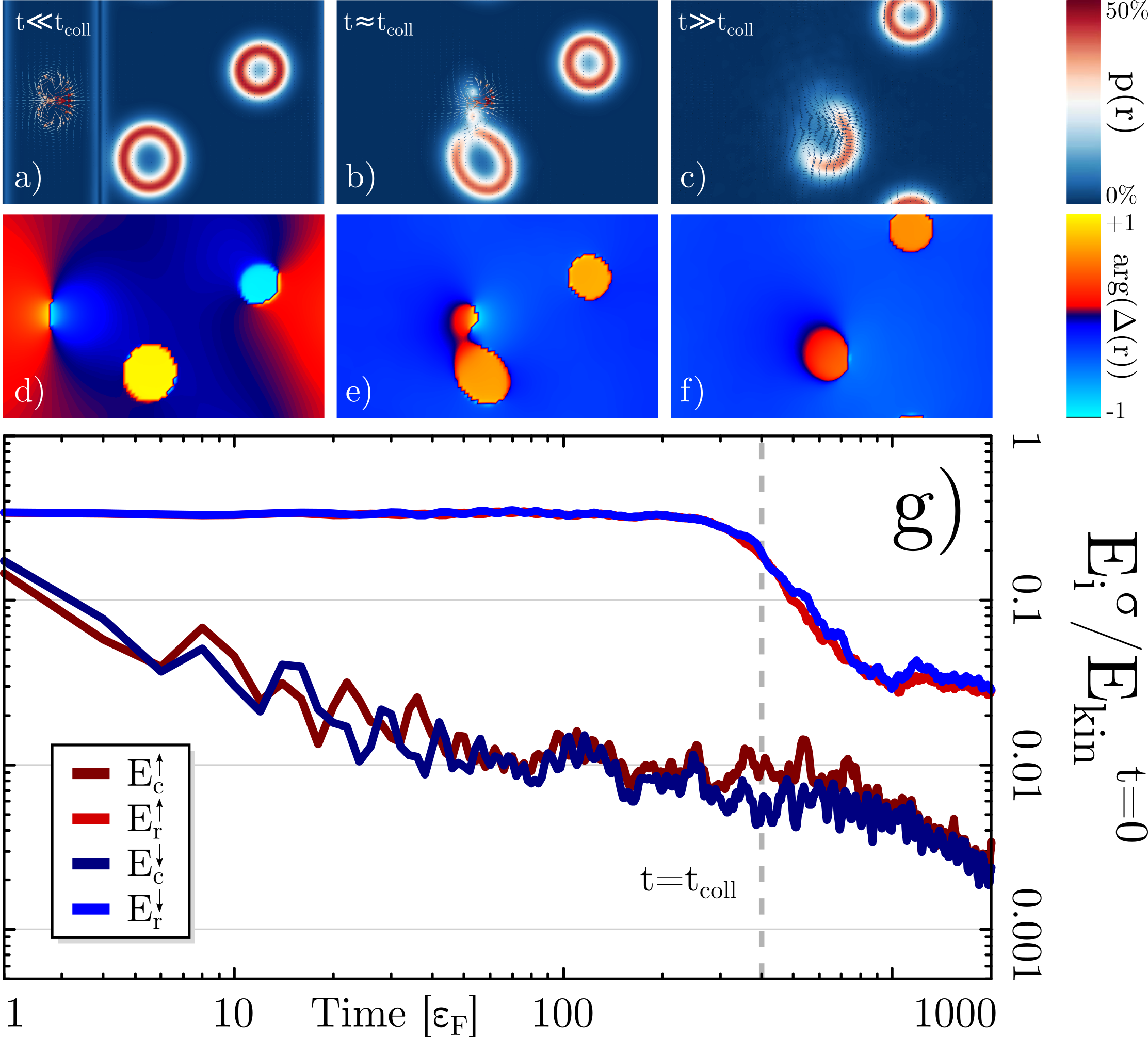}
	\caption{Time evolution of a vortex dipole with initial size $d_{\textrm{i}} = 8k_F^{-1}$ propagating in a system with spin imbalance $P \approx 3.3\%$. Panels (a–c) show maps of the local spin polarization $p(\bm{r})$ at three characteristic times: before the vortex dipole–ferron collision ($t \ll t_\textrm{coll}$), during the collision ($t \approx t_\textrm{coll}$), and after the collision ($t \gg t_\textrm{coll}$). Panels (d–f) display the corresponding phase profiles of the order parameter $\arg\bigl[\Delta(\bm{r})\bigr]$. In this simulation, the dipole annihilates as a result of the collision, while the ferron transforms into a structure that may be viewed either as a deformed ferron or a soliton-like structure. Panel (g) shows the time evolution of the energy components obtained via the Helmholtz decomposition [Eq.~(\ref{eq:pol-hdec})]; the vertical dashed line marks the collision time. \label{fig:lp-split}}
	
\end{figure}

It is instructive to look closer at the cases where the dipole propagation ends with the annihilation process. The example scenario, for $d_{\textrm{i}}=8k_F^{-1}$ and $P\approx 3.3\%$, is visualized in Fig.~\ref{fig:lp-split}a-c. In this configuration, one would expect the dipole with a smaller size (which, in principle, should carry more kinetic energy) to disrupt the ferron through the interaction and annihilation.
This is not observed, and instead, the ferron slowly regenerates and is in motion after the regeneration. We can inspect the phenomena more quantitatively by using the Helmholtz decomposition, which is routinely used to study flow energy in superfluids~\cite{Tsubota2013}. The energy is written in terms of weighted velocity field $\bm{w}=\bm{j}/\sqrt{\rho}=\sqrt{\rho} \bm{v}$, which next is decomposed into compressive ($C$) and rotational ($R$) parts
\begin{equation}
    E_\textrm{kin} = \int \frac{\bm{j}^2(\bm{r})}{2\rho(\bm{r})}d\bm{r}=
    \int \frac{\bm{w}^2(\bm{r})}{2}d\bm{r}=
    \int \frac{\bm{w}_C^2(\bm{r})}{2}d\bm{r}+
    \int \frac{\bm{w}_R^2(\bm{r})}{2}d\bm{r},
\end{equation}
such that $\bm{\nabla}\times \bm{w}_C=0$, $\bm{\nabla}\cdot \bm{w}_R=0$, and $\int \bm{w}_C\cdot \bm{w}_R\, d\bm{r}=0$. The compressive and rotational components are attributed to flows due to sound (phonons) and vortices, respectively.  
In our case, the two energy components are calculated separately for each species:
\begin{align}
	\label{eq:pol-hdec}
	E_\textrm{kin} &= E_\textrm{kin}^\uparrow + E_\textrm{kin}^\downarrow = \int d^3\bm{r} \left( \frac{\bm{j}_\uparrow^2(\bm{r})}{2\rho_\uparrow(\bm{r})} + \frac{\bm{j}_\downarrow^2(\bm{r})}{2\rho_\downarrow(\bm{r})} \right) = \\
	&= (E_C^\uparrow + E_R^\uparrow) + (E_C^\downarrow + E_R^\downarrow). \nonumber
\end{align}
The result of the decomposition is shown in Fig.~\ref{fig:lp-split}g. As long as the dipole does not interact with the ferrons, the rotational components remain fairly constant. The compressive component, in turn, steadily decreases. It is 
typical behavior of dissipative systems: the energy from phonons is transferred to internal degrees of freedom, like quasiparticle excitations. Upon the collision of the vortex dipole with the ferron, we observe a significant decrease in the rotational energy, by one order of magnitude in the discussed case. However, there is no corresponding increase in the compressive component. We only observe that the latter one stabilizes at a given value and starts to decrease further once the annihilation event is completed.

\begin{figure}[t]
	\centering
	\includegraphics[width=0.8\textwidth]{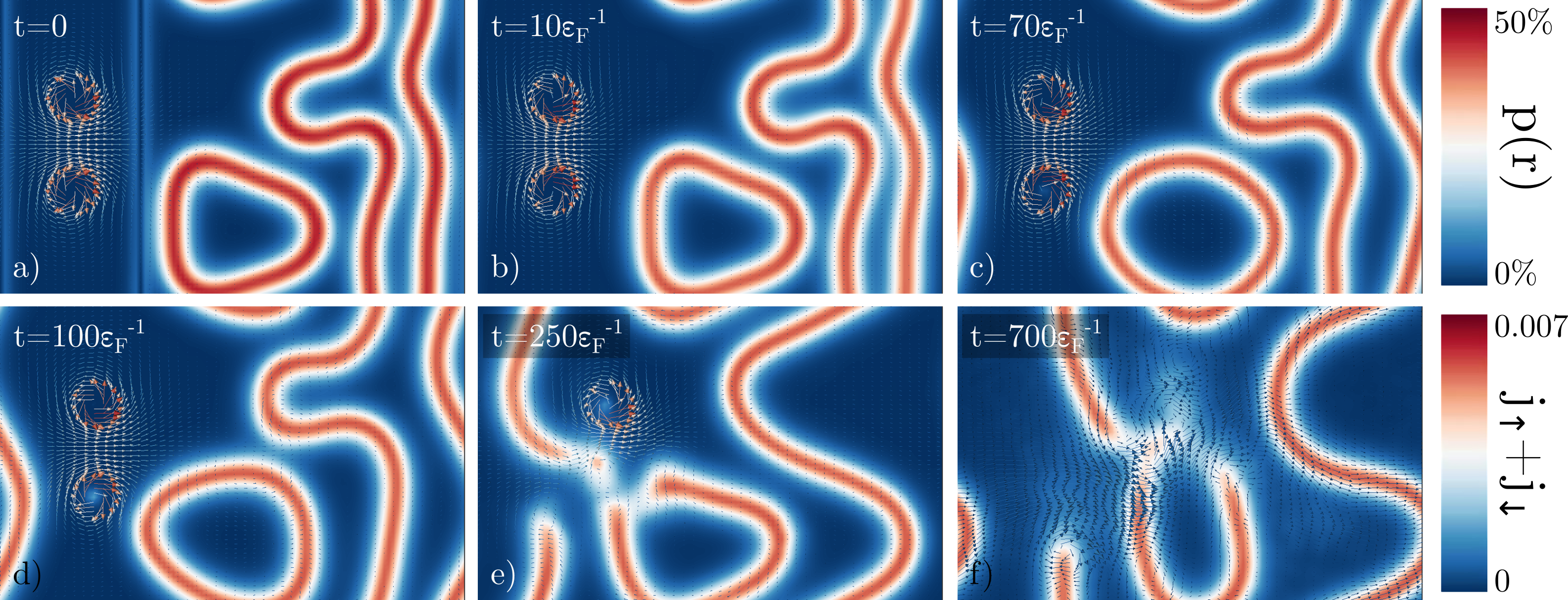}
	\caption{Example of a collision between a vortex dipole of size $d_{\textrm{i}}=20k_F^{-1}$ and a tangle of polarized nodal lines at $P\approx 10\%$. Arrows (heat scale) indicate the local intensity and direction of the total current $\bm{j}_\uparrow(\bm{r})+\bm{j}_\downarrow(\bm{r})$ (in code units); color map (night scale) indicates local polarization $p(\bm{r})$.
		Panel (a): Configuration at time $t=0$, after the dipole has been imprinted. The vertical line at $x=S_x$ indicates the potential split.
		Panel (b): $t \approx 10\eF^{-1}$: the separating potential has completely vanished.
		Panel (c): $t \approx 70\eF^{-1}$: nodal lines start expanding. 
		Panel (d): $t \approx 100\eF^{-1}$: expanding nodal lines collide with the moving dipole; notice the comparison with Panel a): the dipole has moved by a negligible distance, whereas the nodal lines have shifted substantially. Moreover, polarization flows inside the closest vortex core.
		Panel (e): $t \approx 250\eF^{-1}$: the dipole is absorbed by the nodal lines and the circulation starts vanishing.
		Panel (f) $t \approx 700\eF^{-1}$: the dipole has vanished and the nodal lines have occupied the unpolarized portion of the system. The circulation induced by the two vortices has completely dissipated through the system.}
	\label{fig:high-pol}
\end{figure}

The analysis shows that the energy from the dipole is mainly transferred into internal excitations upon interaction with the normal component. This can be understood as follows: in the region where the normal component is concentrated (red rings on the plot with the local polarization $p(\bm{r})$), the energy gap is significantly suppressed ($|\Delta|\approx 0$), which allows to easily excite quasiparticles above the Fermi surface. We also note that the compressive and rotational energies for individual spin components behave in the same way, indicating that energy conversion processes do not differentiate with respect to the spin degree of freedom. 

As a side remark, we also note that trajectories of individual vortices, before and after interaction with the ferron, are not smooth but rather exhibit oscillations, see cases $d_{\textrm{i}}\geq 12\kF^{-1}$ in Fig.~\ref{fig:dfdi-lowpol}b. Such types of oscillations are expected in cases where vortices are massive~\cite{Richaud2021,Dambroise2024,kanjo2024}. Indeed, the applied TDDFT framework admits the massive vortices~\cite{richaud2024}. The mass arises from the normal matter hosted in the vortex core, and since the vortex captures extra matter during the interaction with the ferron, its mass changes as well. It is reflected in a change of frequency and amplitude of the trajectory oscillation after the interaction event. The vortex mass in a spin-imbalanced system is not the main scope of this research, and we do not analyze the properties of trajectory fluctuation in more detail. 
We also rule out the possibility of creation of higher-charged vortices as a result of vortex-ferron interaction, since inspection of the order parameter phase profile shows no change in the winding number.

At higher polarizations, $P \gtrsim 10\%$, the amount of unpaired particles is such that the nodal lines of individual separated ferrons 
cease to exist, and 
the nodal lines 
form a complex pattern dominating the volume portion where no polarization constraints are in place. 
Fig. \ref{fig:high-pol} shows this pattern, reproducing the results of~\cite{ferron-fflo}. Notably, the rate of expansion of the nodal lines 
exceeds the vortex dipole velocity. The latter ends up being absorbed by the nodal lines as soon as the distance between the two is of the order of the coherence length. 
Fig. \ref{fig:high-pol} features the time evolution of a case with $d_{\textrm{i}}=20k_F^{-1}$, but similar qualitative behavior has been observed for a wide range of dipole sizes allowed by our setup. 
Data gathered in this work shows that a vortex dipole, no matter its initial size $d_{\textrm{i}}$, cannot propagate through a highly polarized medium. Upon collision with the normal component localized within nodal lines, the circulation is dispersed, and the vortices vanish. The dissipative forces dominate over the Magnus force in this regime, leading to the annihilation process on a relatively short timescale. 

\section{Summary and Conclusion}

We have explored vortex dipole dynamics in a spin-imbalanced strongly interacting Fermi gas. Intuitively, such systems are expected to exhibit a significant amount of normal component at zero temperature, since not all particles can create Cooper pairs. Indeed, we detect dissipative vortex dipole dynamics, reflected by the decrease of the dipole size as it propagates. From the perspective of the effective vortex point model, this behavior is directly related to the mutual friction between the superfluid and normal components, modeled by the friction forces~(\ref{eq:FN}). The dissipative forces intensify as we increase the spin imbalance, and eventually dominate over the Magnus force~(\ref{eq:FM}). As a consequence, the vortex dipoles can propagate at distances much larger than their size only if the polarization is not too large, $P\lesssim 5\%$; otherwise, they annihilate relatively fast. 

More importantly, we find that dipoles do not propagate along straight trajectories, but they are deflected instead. It demonstrates the existence of a mechanism that differentiates the forces acting on the vortex from those acting on the antivortex. We attribute such symmetry breaking to non-uniformly distributed superfluid $\rho_s(\bm{r})$ and normal $\rho_n(\bm{r})$ components. This interpretation is supported by previous studies such as those presented in Refs.~\cite {ferron-fflo,ferron1,ferron2}, where the emergence of stable structures in spin-imbalanced Fermi gases characterized by spatially varying order parameters has been presented. The observed transverse shift of the dipole, without a clear change in the direction of propagation, points to the transversal force ($\sim D^{\prime}$) as the dominant one. However, this statement cannot be unambiguously verified without detailed modeling of the vortex dynamics with the vortex point model. This, in turn, requires knowledge of the explicit dependence of the dissipative coefficients $D$ and $D^\prime$ as a function of local superfluid and normal densities. These are currently not known. However, we demonstrated that detailed studies of vortex dynamics in spin-imbalanced systems have the potential to deliver useful information for testing various hypotheses, like the presence or absence of the mentioned Iordanskii force.

We expect the results of these simulations to be a useful benchmark for future experiments. We have shown that vortex dipoles can, in fact, be used as probes to examine underlying structures in polarized media in two different ways. Measurements of the dipole size during the propagation quantify the strength of dissipative phenomena driven by the normal component due to spin polarization. 
The shape of the trajectory will, in turn, be a clear indicator of whether the state of the system is described by a uniform or non-uniform distribution of superfluid and normal components. Deflection in the dipole's trajectories should be regarded as a unique consequence of the interaction with a local asymmetry of the normal component. It will, indirectly, point to spatially varying order parameter $\Delta(\bm{r})$, which is the characteristic property of all LOFF-like states or disorder states as postulated in~\cite{ferron-fflo}. 

\ack
This work was financially supported by the (Polish) National Science Center Grants No. 2022/45/B/ST2/00358 (A.B., G.W.) and No. 2021/43/B/ST2/01191 (P.M.).
The majority of this work used computational resources TSUBAME3.0 supercomputer provided by Tokyo Institute of Technology through the HPCI System Research Project (Project IDs: \verb|hp220072|, \verb|hp230081|, \verb|hp240085|).

\section*{Supplementary Data}
Data underlying the study and detailed instructions on how to reproduce the results are available via Zenodo
repository~\cite{zenodo}.

\section*{Notes}
TDDFT calculations and data analysis were performed by AB. PM and GW performed point vortex model analysis. All authors contributed to the interpretation of the results and manuscript writing.

\appendix
\section{Initial states preparation protocol}
\label{Appendix:A}

\begin{figure}[t]
	\centering
	\includegraphics[width=0.6\textwidth]{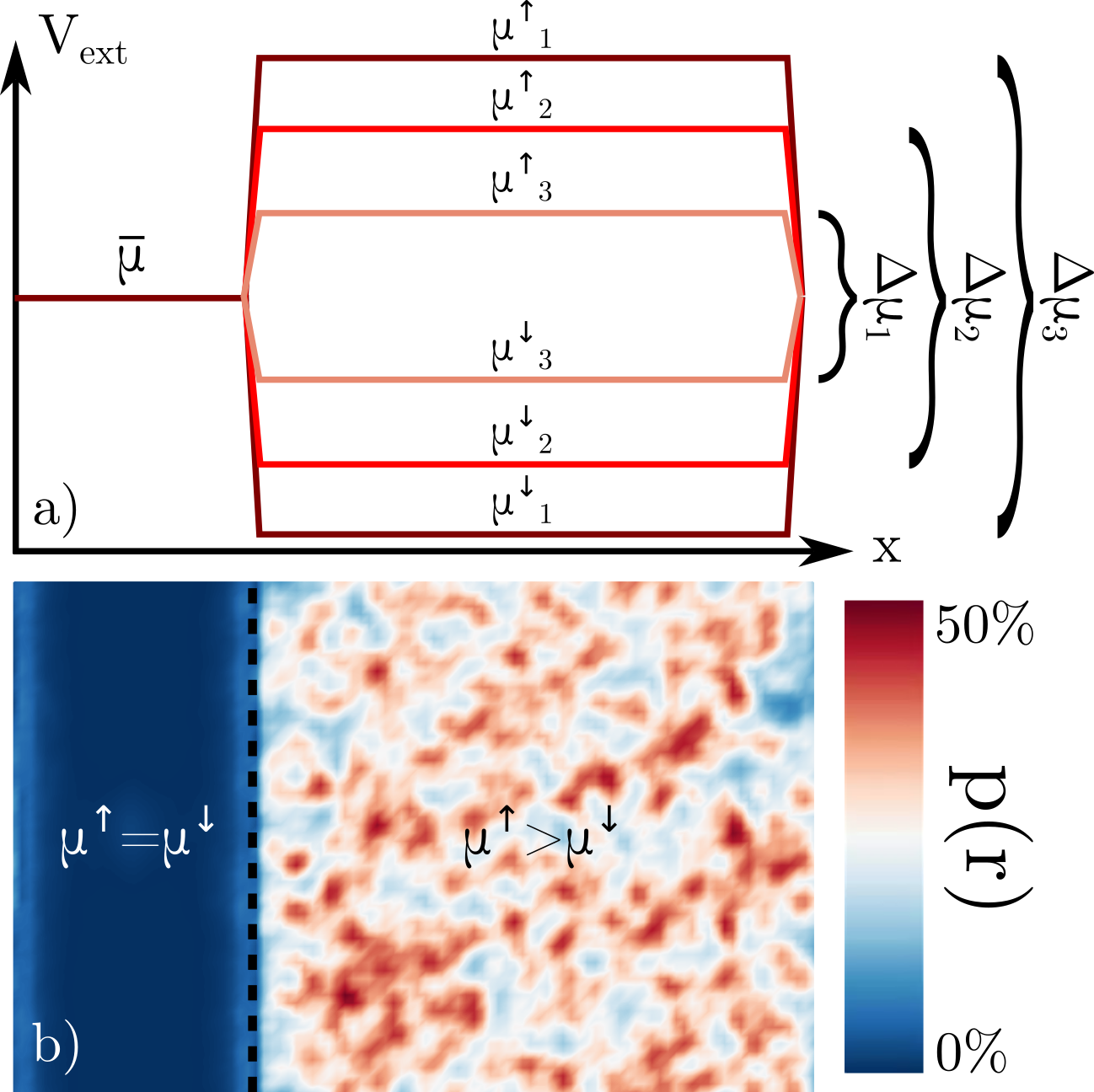}
	\caption{a) Visual representation of the external potential $V^{\textrm{(ext)}}_{\sigma}$ along the x-axis for different values of $\Delta\mu$. b) Polarization across the xy-plane, with additional random density perturbation added to the side with nonzero polarization. }
	\label{fig:mu-split}
\end{figure}
In order to explore the vast landscape of scenarios that vortex motion offers, we need to construct a desirable initial configuration. There are two control parameters that can be tuned independently of each other: initial intervortex distance $d_{\textrm{i}}$, defined as the linear distance between vortex cores, and global spin polarization $P$. 
In order to construct a configuration with the required characteristics, we introduce the spin-dependent external potential $V^{(\textrm{ext})}_\sigma$, that changes local chemical potential $\mu_{\sigma}^{(\textrm{loc})}(\bm{r})=\mu_\sigma - V^{(\textrm{ext})}_\sigma(\bm{r})$ as shown in Fig.~\ref{fig:mu-split}a. The external potential is characterized by two parameters: $S_x$, indicating the coordinate along the x-axis where the chemical potentials start to split, and the splitting value $\Delta \mu$. The changes of the $\mu_{\sigma}^{(\textrm{loc})}(\bm{r})$ around $x=0$ and $x=S_x$ are smoothed, i.e., they are described by the hyperbolic tangent functions. Such a modification to the chemical potentials ensures that we have a spatial separation, so that 
\begin{equation}
	\begin{cases}
		\mu_\uparrow^{(\textrm{loc})} = \mu_\downarrow=\bar{\mu} &  \textrm{for} \,\, x < S_x, \\
		\mu_{\uparrow/\downarrow}^{(\textrm{loc})} =\bar{\mu} \pm \Delta\mu/2  & \textrm{for} \,\, x>S_x.
	\end{cases}
\end{equation}
Furthermore, to generate a ground state that includes one or more ferrons, we introduce a random density perturbation in our system for the $x>S_x$ zone. It is visualized in Fig.~\ref{fig:mu-split}b. By doing this, unpaired particles will cluster together and give rise to ferrons and/or nodal lines, as predicted in Refs~\cite{ferron1, ferron2}. This ensures that the location of ferrons does not systematically impact our analysis.
According to~\cite{ferron-fflo} and reproduced in our configurations with a spatially confined spin polarization, a lower spin imbalance allows for a limited number of relatively large ($R>\xi$) and separated ferrons, while a higher global imbalance gives rise to more compact and spatially extended nodal lines which eventually form complex disordered structures that extend over the whole system.
We highlight that the external potential is activated only during static calculations and is not present during the dynamic runs.

Vortices are imprinted by imposing that the phase of the order parameter in the initial state has a spatial distribution consistent with those expected for a pair of vortex-antivortex. The same protocol was used in work~\cite{barresi}. We imprint the vortex dipole in the region where $x<S_x$, that is, in the unpolarized portion of our simulation domain.
This is to ensure that at the beginning of the simulation, there are no unpaired particles localized inside the cores, and therefore, their dynamic is not impacted. In fact, work~\cite{magierskispin} compares the structure of vortex cores in unpolarized and polarized Fermi gases, which show a fundamentally different vortex core density.
Quantifying the exact relation between global polarization and local polarization inside vortex cores is not necessary for the present study. The phase imprinting procedure is currently deployed in the unpolarized portion to avoid such considerations.
\begin{figure}[t]
	\centering
	\includegraphics[width=0.8\columnwidth]{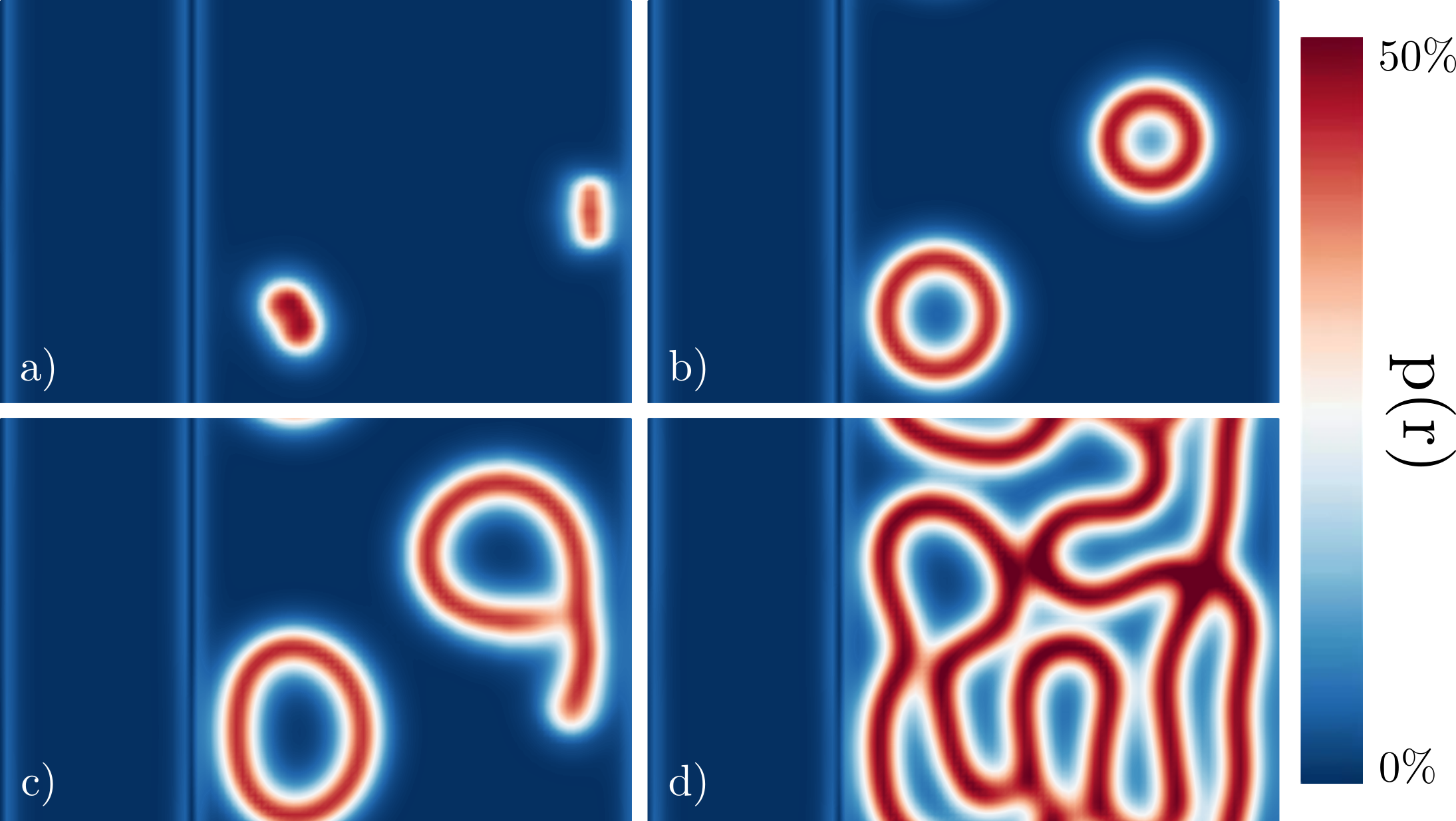}
	\caption{Effect of increasing global polarization on the arising structures in the unitary Fermi gas. The color bar on the right indicates local polarization $p(\bm{r})$. The vertical line at $x=S_x$ is a numerical effect due to the abrupt difference in chemical potentials at its edges. The panels are for spin polarizations $P=$  
		$0.87\%$ (a), $3.30\%$ (b), $6.18\%$ (c) and $17.34\%$ (d).}
	\label{fig:nodal-pol}
\end{figure}

An important additional remark is that, most likely, the global ground state configuration has not been reached in the static part of the simulations. 
The relative energy change between successive iterations, within the self-consistent process, decreased below $|E^{(i+1)}-E^{(i)}|/E^{(i)}\lesssim 10^{-6}$. As noted in Ref.~\cite{ferron-fflo}, this level of accuracy may still be insufficient to guarantee access to the absolute ground state, since the energy landscape of spin-imbalanced systems can exhibit many local minima with energy differences comparable to this threshold.
The produced initial states, therefore, are attributed to some amount of the excitation energy.
However, the phenomena that we seek to examine do not strictly require the state to be in the global minimum of energy; in fact, the addition of a vortex dipole already adds enough excitations to push any ground state out of its local minimum. 
The lack of absolute convergence does not physically alter the system under study. Fig.~\ref{fig:nodal-pol} shows examples of initial states used in this study before applying the vortex dipole imprint procedure. Specifically, a comparatively small change in chemical potential difference $\Delta \mu$ can cause a large increase in polarization, therefore bringing the system from configurations with clearly separated ferrons (panels a,b) to others with more densely packed nodal lines where ferron boundaries are not clear (panels c,d).

\section*{References}
\providecommand{\newblock}{}

\end{document}